\documentclass[a4paper, 10pt, fleqn]{scrartcl}
\usepackage[british]{babel}
\usepackage{amsfonts,amssymb,microtype,mathdefs,array,mathrsfs,amsthm,enum2,relsize}
\usepackage[section]{thmdefs2}
\usepackage[hidelinks]{hyperref}

\setkomafont{pagehead}{\small\itshape}
\setkomafont{sectioning}{\rmfamily\bfseries}
\setkomafont{descriptionlabel}{\itshape}

\typearea{calc}

\makeatletter
\newcommand\m@thsm@ller[2]{\mbox{\relscale{0.91}$\m@th#1#2$}}
\let\smaller\undefined
\DeclareRobustCommand\smaller[1]{\relax\ifmmode{\mathpalette\m@thsm@ller{#1}}\else{\relscale{0.91}#1}\fi}
\makeatother

\DeclareRobustCommand*{\dom}{\qopname\relax o{dom}}

\DeclareRobustCommand*{\supp}{\qopname\relax o{supp}}
\newcommand*{\id}{\mathrm{id}}

\newcommand*{\Set}{\mathsf{Set}}
\newcommand*{\Alg}{\mathrm{Alg}}
\newcommand*{\Pol}{\mathrm{Pol}}

\newcommand*{\Flat}{\mathrm{flat}}
\newcommand*{\sing}{\mathrm{sing}}

\newcommand*{\suc}{\mathrm{suc}}
\newcommand*{\act}{\mathrm{act}}

\newcommand*{\Min}{\mathrm{Min}}

\newcommand*{\SG}{\mathrm{SG}}

\newcommand*{\FO}{\smaller{\mathrm{FO}}}

\newcommand*{\Syn}{\mathrm{Syn}}

\newcommand*{\emptyseq}{\smaller{\langle\rangle}}

\newcommand*{\?}{\kern .08em}

\newcommand*{\lsem}{[\![}
\newcommand*{\rsem}{]\!]}

\newcommand\upqed{\vskip-\baselineskip\vskip-\belowdisplayskip}

\begin{document}
\title{Some Remarks on First-Order Definable Tree Languages}
\author{Achim Blumensath}
\maketitle

\begin{abstract}
We study the question of whether a given regular language of finite trees can
be defined in first-order logic. We develop an algebraic approach to address
this question and we use it to derive several necessary and sufficient
conditions for definability (but unfortunately no condition that is both).
The main difference of our results to those from the literature is that our conditions
are decidable.
\end{abstract}

\section{Introduction}   

The question of how to decide whether a given language of trees is definable in first-order
logic is a long-standing open problem in language theory.
It has first been raised in \cite{Thomas84}, 40 years ago.
Since then the problem has withstood numerous attempts to solve it.

Benedikt and Segoufin~\cite{BenediktSegoufin09} have provided a decidable characterisation
for languages that are definable using only the successor relation, but not the tree order.
The proof is mostly model-theoretic and it makes essential use of locality arguments,
which do not work in the presence of the tree order.

There also exist several non-effective characterisations for languages definable with
the tree order. The seminal article by Thomas~\cite{Thomas84} contains a mixed characterisation
in terms of a temporal logic. A~related result can be found in~\cite{HaferThomas87}.
A~characterisation in terms of star-free expressions has been provided by Heuter~\cite{Heuter91}.
There are also three publications with algebraic characterisations
by Esik and Weil~\cite{EsikWeil10}, Boja\'nczyk~\cite{Bojanczyk04},
and Boja\'nczyk, Straubing, and Walukiewicz~\cite{BojanczykStWa12}.

Finally, there are some partial results. The papers~\cite{Heuter91,Potthoff92,Potthoff95}
by, respectively, Heuter and Potthoff, provide counterexamples showing, among other things,
that `aperiodicity' is insufficient as a criterion for definability. Let us also mention
a paper by Boja\'nczyk and Michalewski~\cite{BojanczykMiXX} that contains several observations
regarding first-order definability, although its main focus is on a different logic.

The main reason why the definability question has turned out to be that resistant to attacks
seems to be that we have not yet developed the right combinatorical and algebraic tools
to resolve the question.
Boja\'nczyk and Michalewski~\cite{BojanczykMiXX} suggest to use the algebraic machinery
from Tame Congruence Theory~\cite{HobbyMcKenzie88}, a subfield of Universal Algebra,
to attack the definability problem. Other possible tools include analogues to
Simon Factorisation Trees~\cite{BlumensathYY}.
But both approaches do not seem to be strong enough to fully settle the question.

The last two decades saw the development of several algebraic theories for tree languages
(see~\cite{Bojanczyk21} for an overview).
The two most developed ones of these seem to be the one based on
forest algebras introduced in~\cite{BojanczykWalukiewicz07} and
the monadic framework from~\cite{Blumensath20}.

In this article we follow~\cite{BojanczykMiXX} and try to apply techniques from Tame Congruence
Theory~\cite{HobbyMcKenzie88} to the definability problem for first-order logic.
We are able to derive a (decidable) necessary condition for first-order in
Corollary~\ref{Cor: neccessary condition} below, and two (decidable) sufficient ones in
Corollary~\ref{Cor: sufficient condition 1} and
Proposition~\ref{Prop: conditions for FO-definability}.
While non-trivial, most of our results are not very deep.

Our motivation for this work stems from the fact that,
when setting up an algebraic framework to address definability questions for tree languages,
there are many choices to make regarding the technical details.
While these choices seem to be arbitrary at first,
they turn out to matter a lot for how well the framework functions.
Examples of such choices include \textsc{(i)}~whether to work with linear or with non-linear
trees (we have to use linear trees for first-order logic)\?;
\textsc{(ii)}~whether to work with simple contexts or with multi-contexts
(we need multi-contexts)\?;
\textsc{(iii)}~which notion of a subalgebra one uses
(we need subalgebras where we can remove both elements and functions)\?; etc.
Having worked out all the technical details, I~therefore consider it worthwhile to
share my results with the public so that others will not have to repeat the same work
and to provide a basis from which to attack the first-order definability problem for real.

The overview of the article is as follows. We start in Section~\ref{Sect:tree algebras}
with setting up the algebraic framework we will be working in.
In Section~\ref{Sect: survey}, we give a short overview of known results from the
literature. Our results make use of tools from Tame Congruence Theory
which we present in Section~\ref{Sect:minimal}.
Before doing so we have to provide a translation between preclones (the algebras we are
working with) and clones (the kind of algebras used in Tame Congruence Theory).
This is done in Section~\ref{Sect:congruences}.
The two final sections contain our results about first-order definability.
Section~\ref{Sect:non-definable} derives necessary conditions, while
Section~\ref{Sect:definable} contains sufficient ones.

\section{Tree Algebras}   
\label{Sect:tree algebras}

We start by introducing the algebraic framework used in this article,
which is a special case of that from~\cite{Blumensath20,Blumensath21}.
To make the paper more accessible to a general audience we have tried to keep
the category-theoretical prerequisites at a minimum.

We will be working with two different kinds of algebraic structures\?: \emph{clones} and
\emph{preclones.} Both are structures where all operations are treated as elements,
that is, a (pre-)clone is a set of operations (of various arities) that can be composed.
The difference is whether or not the terms resulting form these
compositions can use variables several times or only once.
To defines this formally, we fix some countably infinite set~$X$ of \emph{variables.}
Let $\Xi := \PSet_\omega(X)$ be the set of all finite sets of variables.
We consider \emph{$\Xi$-sorted sets}~$A$ where each element has an associated \emph{arity}
$\xi \in \Xi$. Such elements are interpreted as `operations using the variables in~$\xi$'.
Then $A = (A_\xi)_{\xi \in \Xi}$ is a family of sets where $A_\xi$~denotes
the subset of elements of arity~$\xi$. A \emph{function} $f : A \to B$ between $\Xi$-sorted
sets is a family $f = (f_\xi)_{\xi \in \Xi}$ of functions $f_\xi : A_\xi \to B_\xi$.
In category-theoretical language this means that we work inside the category $\Set^\Xi$
of $\Xi$-sorted sets.
To keep terminology simple, we will usually use the terms \emph{set} and \emph{function}
dropping the adjective \emph{$\Xi$-sorted.}
\begin{Def}
Let $A$~be a ($\Xi$-sorted) set.

(a) We set
\begin{align*}
  \bbC A := (\bbC_\xi A)_{\xi \in \Xi}\,,
\end{align*}
where $\bbC_\xi A$ denotes the set of all finite terms~$t$ with operations from~$A$
and variables from~$\xi$ such that that every variable occurs at least once in~$t$.

(b) Similarly, we set
\begin{align*}
  \bbT A := (\bbT_\xi A)_{\xi \in \Xi}\,,
\end{align*}
where $\bbT_\xi A$ is the set of all $t \in \bbC_\xi A$ such that
\begin{itemize}
\item every variable $x \in \xi$ appears exactly once in~$t$ and
\item the whole term~$t$ does not consist of just a variable.
\end{itemize}
We call the element of~$\bbT A$ \emph{linear trees,} those of~$\bbC A$ \emph{non-linear} ones.

(c) For a function $f : A \to B$, we denote by $\bbC f : \bbC A \to \bbC B$ and
$\bbT f : \bbT A \to \bbT B$ the functions applying~$f$ to each operation in
the given term. (The variables are left unchanged.)
\end{Def}
In the following we will frequently treat terms $t \in \bbC_\xi A$ as labelled trees of the form
$t : \dom(t) \to A + \xi$, where $\dom(t)$~denotes the set of vertices of~$t$.
This allows us to use terminology from terms and trees interchangeably.
\begin{Def}
Let $t \in \bbC_\xi A$.

(a) Let $x$~be a variable. We call a vertex~$v$ of~$t$ an \emph{$x$-successor} of a vertex~$u$
if it is the successor of~$u$ that corresponds to the argument of the operation~$t(u)$
for the variable~$x$.

(b) We denote the subtree of~$t$ \emph{attached} at a vertex~$v$ by~$t|_v$.
\end{Def}

There are two natural operations on the above sets of terms.
\begin{Def}
Let $A$~be a set.

(a) The function $\Flat : \bbC\bbC A \to \bbC A$ takes a term where each operation
is itself a term over~$A$ and substitutes each of the small terms into its parent. That is,
inductively,
\begin{align*}
  \Flat(s(t_0,\dots,t_{n-1})) :=
    s\bigl(\Flat(t_0),\dots,\Flat(t_{n-1})\bigr)\,,
\end{align*}
for $s \in \bbC A$ and $t_i \in \bbC\bbC A$.

(b) The operation $\sing : A \to \bbC A$ maps each element $a \in A$ to the corresponding
term consisting just of~$a$ and variables. That is,
\begin{align*}
  \sing(a) := a(x_0,\dots,x_{n-1})\,,
  \quad\text{for } a \in A_\xi \text{ with } \xi = \{x_0,\dots,x_{n-1}\}\,.
\end{align*}

(c) We denote the corresponding restrictions $\bbT\bbT A \to \bbT A$ and $A \to \bbT A$
to~$\bbT$ by the same names $\Flat$ and $\sing$.
\end{Def}
\begin{Exam}
Given the terms
\begin{alignat*}{-1}
  s(x,y,z) &:= a(a(x,y), b(z))\,, \qquad
  &u(x)     &:= a(c, x)\,, \\
  t(x,y)   &:= b(a(x,y))\,, \qquad
  &v(x)     &:= b(a(x,x))\,,
\end{alignat*}
we have
\begin{align*}
  &\Flat\bigl(s(t(y,x), u(z), v(y))\bigr) \\
  &\quad{}= a\bigl(a\bigl(b(a(y,x)),\, a(c,z)\bigr),\,
                   b\bigl(b(a(y,y))\bigr)\bigr)\,.
\end{align*}
\upqed
\end{Exam}

Having introduced two different kinds of terms, we can define the type of algebras
we are interested in as follows.
\begin{Def}
Let $\bbM$~be one of $\bbC$~or~$\bbT$.

(a) An \emph{$\bbM$-algebra} is a pair $\frakA = \langle A,\pi\rangle$ consisting of a set~$A$
and a \emph{product} $\pi : \bbM A \to A$ satisfying the axioms
\begin{align*}
  \pi \circ \sing = \id
  \qtextq{and}
  \pi \circ \bbM\pi = \pi \circ \Flat\,.
\end{align*}
The first one is called the \emph{unit law,} and the second one the \emph{associative law.}
$\bbC$-algebras are also called \emph{clones} and $\bbT$-algebras are \emph{preclones.}

(b) A \emph{morphism} $\varphi : \frakA \to \frakB$ of $\bbM$-algebras is a function
$\varphi : A \to B$ commuting with the respective products in the sense that
\begin{align*}
  \pi \circ \bbM\varphi = \varphi \circ \pi\,.
\end{align*}
\upqed
\end{Def}
\begin{Rem}
For readers familiar with category-theoretic terminology, note that the triples
$\langle\bbC,\Flat,\sing\rangle$ and $\langle\bbT,\Flat,\sing\rangle$ form monads on $\Set^\Xi$,
and that our notion of an algebra is that of an Eilenberg-Moore algebra for the corresponding
monad.
\end{Rem}

\begin{Exams}
(a) Given an algebraic signature~$\Sigma$,
we can associate the so-called \emph{polynomial clone}~$\Pol(\frakA)$ with
a $\Sigma$-algebra~$\frakA$ as follows.
The domain~$\Pol_\emptyset(\frakA)$ consists of all elements of~$A$ while $\Pol_\xi(\frakA)$,
for $\xi \neq \emptyset$, consists of all functions $f : A^\xi \to A$ of the form
\begin{align*}
  f(\bar a) := s(\bar a,\bar c)\,,
\end{align*}
for some $\Sigma$-term $s(\bar x,\bar y)$ and some tuple~$\bar c$ of elements of~$A$.
The product $\pi(t)$ of a term $t \in \bbC\Pol(\frakA)$ is just the composition of functions.

Conversely, we can associate with every clone~$\frakC$ an algebra~$\Alg(\frakC)$
over the signature $\Sigma := C$ as follows.
For the universe we take the set~$C_\emptyset$ consisting of all elements of
arity~$\emptyset$.
For every $c \in C_\xi$, we add one operation $c : C_\emptyset^\xi \to C_\emptyset$
mapping a tuple~$\bar a$ to the product $\pi(s)$, where $s$~is the term $c(\bar a)$.

(b) For every set~$A$, the pair $\langle\bbM A,\Flat\rangle$ forms an $\bbM$-algebra
called the \emph{free} $\bbM$-algebra over~$A$.

(c) Finally, with every $\bbM$-algebra~$\frakA$ we can associate a semigroup
$\SG(\frakA)$ consisting of all elements of~$A$ of sort~$\{z\}$
(for some arbitrary but fixed variable~$z$). The product of~$\SG(\frakA)$
is defined by
\begin{align*}
  ab := \pi\bigl(a(b(z))\bigr)\,, \quad\text{for } a,b \in A_{\{z\}}\,,
\end{align*}
where the product on the right-hand side is computed in~$\frakA$.
\end{Exams}

\begin{Rem}
Every $\bbC$-algebra~$\frakA = \langle A,\pi\rangle$ has an associated $\bbT$-algebra
$\frakA|_\bbT := \langle A,\pi_0\rangle$ where $\pi_0 : \bbT A \to A$ is the restriction
of $\pi : \bbC A \to A$.
\end{Rem}

We will frequently consider elements $a \in A_\xi$ of an $\bbM$-algebra as operations
$A_\emptyset^\xi \to A_\emptyset$, and elements $t \in \bbM A$ as terms over these operations.
\begin{Def}
Let $\frakA$~be a $\bbT$-algebra or a $\bbC$-algebra.
For $a \in A_\xi$ and $\bar c \in A^\xi$, we set
\begin{align*}
  a(\bar c) := \pi(t)\,,
\end{align*}
where the term~$t$ is obtained from applying the operation~$a$ to the terms~$\sing(c_x)$,
for every $x \in \xi$.
Hence, if $c_x$~has sort~$\eta_x$, then the sort of~$a(\bar c)$ is $\bigcup_{x \in \xi} \eta_x$.
\end{Def}

%
%
%

We will use $\bbT$-algebras to recognise languages of finite trees.
\begin{Def}
Let $\Sigma$~be a finite set.

(a)
A language $K \subseteq \bbT_\xi\Sigma$ is \emph{recognised} by a $\bbT$-algebra~$\frakA$
if there exists a morphism $\varphi : \bbT\Sigma \to \frakA$ and a set $P \subseteq A_\xi$
such that
\begin{align*}
  K = \varphi^{-1}[P]\,.
\end{align*}

(b) We denote first-order logic by $\FO$.
When using $\FO$-formulae to define subsets of $\bbC\Sigma$ or $\bbT\Sigma$,
we represent a tree $t \in \bbC_\xi\Sigma$ as a relational structure of the form
\begin{align*}
  \langle\dom(t),{\leq},(\suc_x)_x, (P_c)_{c \in \Sigma + \xi}\rangle
\end{align*}
where $\leq$~is the tree order (with the root as minimal element),
$\suc_x$~is the $x$-th successor relation (connecting ever vertex with
the successor corresponding to the argument for the variable~$x$),
and $P_c$~is the set of vertices labelled by the element $c \in \Sigma + \xi$.
\end{Def}

We can characterise a class of languages by describing the class of algebras recognising
them. For our purposes, we are interested in the following two properties.
\begin{Def}
Let $\frakA$~be an $\bbT$-algebra.

(a) $\frakA$~is \emph{finitary} if it is finitely generated and
$A_\xi$~is finite, for every sort $\xi \in \Xi$.

(b) $\frakA$~is \emph{$\FO$-definable} if there exists a finite set $C \subseteq A$
of generators and, for every $a \in A$, there is an $\FO$-formula~$\varphi$ such that
\begin{align*}
  \pi(t) = a \quad\iff\quad t \models \varphi\,,
  \quad\text{for every } t \in \bbT C\,.
\end{align*}
\upqed
\end{Def}
\begin{Thm}[\cite{Blumensath21}]
Let $\Sigma$~be a finite set and $K \subseteq \bbT_\xi\Sigma$.
\begin{enuma}
\item $K$~is regular if, and only if, it is recognised by a finitary $\bbT$-algebra.
\item $K$~is $\FO$-definable if, and only if, it is recognised by an $\FO$-definable
  $\bbT$-algebra.
\end{enuma}
\end{Thm}
\begin{Rem}
The second statement is not true for $\bbC$-algebras.
This is why we have to work with $\bbT$-algebras below.
\end{Rem}

In general, there can be many algebras recognising a given language~$K$.
But there always is a minimal one. It is called the \emph{syntactic algebra}
$\Syn(K)$ of~$K$.
\begin{Def}
Let $K \subseteq \bbT_\xi\Sigma$. A $\bbT$-algebra~$\frakA$ is the \emph{syntactic algebra}
of~$K$ if there exists a morphism $\eta : \bbT\Sigma \to \frakA$ recognising~$K$
such that, for every surjective morphism $\varphi : \bbT\Sigma \to \frakB$ recognising~$K$
there exists a unique morphism $\rho : \frakB \to \frakA$ with $\eta = \rho \circ \varphi$.
\end{Def}

The following result is folklore. It seems to have first been observed in~\cite{Thomas84}.
A~detailed proof in the terminology of this article can be found in~\cite{Blumensath21}.
\begin{Thm}
Every regular language $K \subseteq \bbT_\xi\Sigma$ has a syntactic algebra~$\Syn(K)$.
\end{Thm}

We can use syntactic algebras to study $\FO$-definability because of the following observation.
\begin{Thm}[\cite{Blumensath21}]\label{Thm: FO-definable if Syn(K) FO-definable}
A language $K \subseteq \bbT_\xi\Sigma$ is $\FO$-definable if, and only if,
its syntactic algebra $\Syn(K)$ is $\FO$-definable.
\end{Thm}

\section{Existing results}   
\label{Sect: survey}

To put this article into context,
let us briefly mention some of the known results about first-order definable tree languages
from the literature. We start with a result explaining why we have to work with~$\bbT$
instead of~$\bbC$\?: languages of non-linear trees are not closed under inverse homomorphisms.
\begin{Prop}[Potthoff~\cite{Potthoff95}]
There exist finite alphabets $\Sigma$~and~$\Gamma$, a morphism
$\varphi : \bbC\Sigma \to \bbC\Gamma$, and
an $\FO$-definable language $K \subseteq \bbC_\xi\Gamma$ such that the preimage
$\varphi^{-1}[K] \subseteq \bbC_\xi\Sigma$ is not $\FO$-definable.
\end{Prop}

As already mentioned above, several non-effective characterisations are known
of when a tree language is $\FO$-definable.
We start with a characterisation in terms of star-free expressions.
\begin{Def}
Let $\Sigma$~be an alphabet and $\xi \in \Xi$ a sort.
A \emph{star-free expression} over~$\Sigma$ of sort~$\xi$
is a finite term~$\alpha$ built up from the following operations\?:
\begin{itemize}
\item the empty set $\emptyset$,
\item single letters $a(x_0,\dots,x_{n-1})$, where $a \in \Sigma$ is a letter of arity~$n$
  and $x_0,\dots,x_{n-1}$ is an enumeration (without repetitions) of~$\xi$,
\item union~$+$ and complement~$\sim$,
\item concatenation $\alpha \cdot_z \beta$ where $\alpha$~is an expression of sort
  $\eta + \{z\}$, $\beta$~one of sort~$\zeta$, and we have $\eta \cup \zeta = \xi$ and
  $\eta \cap \zeta = \emptyset$.
\end{itemize}
The value $\lsem\alpha\rsem \subseteq \bbT_\xi\Sigma$ of an expression~$\alpha$ of sort~$\xi$
is defined inductively as follows.
\begin{alignat*}{-1}
  \lsem\emptyset\rsem      &:= \emptyset\,, \qquad&
  \lsem\alpha + \beta\rsem &:= \lsem\alpha\rsem \cup \lsem\beta\rsem\,, \\
  \lsem a(x_0,\dots,x_{n-1}) \rsem &:= \{a(x_0,\dots,x_{n-1})\}\,, \qquad&
  \lsem{\sim}\alpha\rsem   &:= \bbT_\xi\Sigma \setminus \lsem\alpha\rsem\,,
\end{alignat*}
and $\lsem\alpha \cdot_z \beta\rsem$ is the set of all trees obtained from some tree
$s \in \lsem\alpha\rsem$ by replacing the leaf with label~$z$ by some tree
$t \in \lsem\beta\rsem$.
\end{Def}
\begin{Thm}[Heuter~\cite{Heuter91}]
A language $K \subseteq \bbT_\xi\Sigma$ is\/ $\FO$-definable if, and only if,
it is the value of some star-free expression.
\end{Thm}
Note that this result is not true when working with star-free expressions based on trees
in~$\bbC\Sigma$, i.e., expressions where we allow a variable to appear several times.
In fact, we have the following result for such expressions.
\begin{Thm}[Potthoff, Thomas~\cite{PotthoffThomas93}]
Let $\Sigma$~be an alphabet without unary letters.
A~language $K \subseteq \bbC_\xi\Sigma$ is is the value of some star-free expression over~$\bbC$
if, and only if, it is regular.
\end{Thm}

The characterisation of $\FO$-definable word languages in terms of aperiodic semigroups
generalises only partially to tree languages.
\begin{Def}
(a) A semigroup~$\frakS$ is \emph{aperiodic} if it has no subalgebra that forms a group.

(b) \emph{Chain logic} is the semantic fragment of monadic second-order logic
where quantification is restricted to sets that form chains with respect to the tree ordering.
\end{Def}
\begin{Rem}
Equivalently, we can define $\frakS$~to be aperiodic if there exists some constant $n < \omega$
such that $a^n = a^{n+1}$, for all $a \in S$.
\end{Rem}
\begin{Thm}[Thomas~\cite{Thomas84}]\label{Thm: FO = chain + aperiodic}
A language $K \subseteq \bbT_\xi\Sigma$ is\/ $\FO$-definable if, and only if,
$K$~is definable in chain logic and\/ $\SG(\Syn(K))$ is aperiodic.
\end{Thm}
It is decidable whether $\SG(\Syn(K))$ is aperiodic, but the decidability of definability in
chain logic is still open. Furthermore, it is known that aperiodicity alone is not sufficient.
\begin{Prop}[Heuter~\cite{Heuter91}]
There exists a language $K \subseteq \bbT_\xi\Sigma$ such that\/ $\SG(\Syn(K))$ is aperiodic,
but $K$~is not\/ $\FO$-definable.
\end{Prop}

Finally, there are three purely algebraic characterisations that are based on a generalisation
of the notions of a block product and a wreath product from semigroup theory.
To keep this section short we omit the definitions of the algebras and operations
in question, hoping to convey at least the flavour of the statements.
The interested reader is referred to the original papers.
\begin{Thm}[Boja\'nczyk, Straubing, Walukiewicz~\cite{BojanczykStWa12}]
A language $K \subseteq \bbT_\xi\Sigma$ is\/ $\FO$-definable if, and only if,
it is recognised by an iterated wreath product of aperiodic path algebras.
\end{Thm}
\begin{Thm}[Boja\'nczyk, Michalewski~\cite{BojanczykMiXX}]
A language $K \subseteq \bbT_\xi\Sigma$ is\/ $\FO$-definable if, and only if,
$\SG(\Syn(K))$ is aperiodic and $K$~is recognised by an iterated wreath product of
matrix powers of the $2$-element semilattice.
\end{Thm}
\begin{Thm}[\'Esik, Weil~\cite{EsikWeil10}]
A language $K \subseteq \bbT_\xi\Sigma$ is\/ $\FO$-definable if, and only if,
$\Syn(K)$ belongs to the smallest pseudo-variety that contains a certain
$\bbT$-algebra~$\frakT_\exists$ and that is closed under block products.
\end{Thm}
The problem with making these characterisations effective is the we do not know how
to compute the number of iterations of the wreath product or block product needed.

\section{Congruences}   
\label{Sect:congruences}

The theorems from the end of Section~\ref{Sect:tree algebras} allow us to use
algebraic techniques to study $\FO$-definable languages.
Unfortunately, the algebraic tools we will use below are formulated for $\bbC$-algebras,
while our results concerning $\FO$-definable languages require $\bbT$-algebras.
We therefore start by presenting a way to translate between these two types of algebras.
Going from a $\bbC$-algebra to a $\bbT$-algebra is trivial\?: we just have to restrict
the product.
For the converse, we have to add elements like $a(x,x)$ that contain some variable~$x$
several times. We do so by taking elements of the form $\langle\sigma,a\rangle$
where $a$~is an element of the given $\bbT$-algebra and $\sigma$~is a piece of additional
information telling us which of the variables of~$a$ have to be identified.
\begin{Def}
(a) Let $\frakA$~be a $\bbC$-algebra. For an element $a \in A_\xi$ and a surjective function
$\sigma : \zeta \to \xi$, we define
\begin{align*}
  {}^\sigma a := \pi\bigl(a(x_{\sigma(z_0)},\dots,x_{\sigma(z_{n-1})})\bigr)\,,
  \quad\text{where } \zeta = \{z_0,\dots,z_{n-1}\}\,.
\end{align*}

(b)
For a set~$A$, we define
\begin{align*}
  \bbX A := (\bbX_\xi A)_{\xi \in \Xi}\,,
\end{align*}
where
\begin{align*}
  \bbX_\xi A :=
    \xi + 
    \bigset{ \langle \sigma,a\rangle }
           { a \in A_\zeta,\ \sigma : \zeta \to \xi \text{ surjective} }\,.
\end{align*}
For a function $f : A \to B$, we define $\bbX f : \bbX A \to \bbX B$ by
\begin{align*}
  \bbX f(\langle\sigma,a\rangle) := \langle\sigma,f(a)\rangle\,.
\end{align*}

(c) For a $\bbT$-algebra $\frakA = \langle A,\pi\rangle$, we set
\begin{align*}
  \bbX\frakA := \langle \bbX A,\hat\pi\rangle\,,
\end{align*}
where the product $\hat\pi : \bbC\bbX A \to \bbX A$ is defined as follows.
Given $t \in \bbC_\xi\bbX A$,
let $G$~be the multi-graph with vertices $\dom(t)$ and the following edges.
There is an $x$-labelled edge $u \to v$ if, and only if, $u$~has label $\langle\sigma,a\rangle$
in~$t$ and $v$~is the $z$-successor of~$u$ in~$t$ with $\sigma(z) = x$.
(There is one such edge for every $z \in \sigma^{-1}(x)$.)
Furthermore, there is an unlabelled edge $u \to v$ if, and only if,
$u$~is labelled by a variable~$x$ and $v$~is the $x$-successor of~$u$ in~$t$.
Let $s$~be the tree obtained from the unravelling of~$G$ by
\begin{itemize}
\item contracting all unlabelled edges,
\item replacing all variables by new ones such that every variable appears exactly once in~$s$.
\end{itemize}
Let $\tau$~be the function mapping each new variable in~$s$ to the corresponding old one.
We set
\begin{align*}
  \hat\pi(t) := \begin{cases}
                  \tau(x)                   &\text{if } s = x \text{ is a variable,} \\
                  \langle\tau,\pi(s)\rangle &\text{otherwise.}
                \end{cases}
\end{align*}
\upqed
\end{Def}
We need the following properties of the operation~$\bbX$ (further details can be found
in~\cite{Blumensath23}).
\begin{Prop}
Let\/ $\frakA$~be a $\bbT$-algebra.
\begin{enuma}
\item $\bbX\frakA$~is a $\bbC$-algebra.
\item There exists an embedding $i : \frakA \to (\bbX\frakA)|_{\bbT}$
  which is bijective on elements of sort~$\emptyset$.
\end{enuma}
\end{Prop}
\begin{proof}
(a) The proof is straightforward but a bit tedious.
The central argument is that, given a graph~$G$ that is partitioned into several subgraphs,
the unravelling of~$G$ can be obtained by first unravelling all the subgraphs
and then the resulting graph.
We omit the details, which can be found in~\cite{Blumensath23}.

(b) We set $i(a) := \langle\id, a\rangle$.
This function is a morphism of $\bbT$-algebras since the way the product
of $\bbX A$ is defined we have
\begin{align*}
  \pi(\bbT i(t)) = i(\pi(t))\,, \quad\text{for all } t \in \bbT A\,.
\end{align*}
\upqed
\end{proof}
\begin{Rem}
Using this result, we can translate every $\bbT$-algebra~$\frakA$ into a
$\bbC$-algebra~$\bbX\frakA$. Unfortunately, the resulting algebra is usually not finitary.
Therefore, we modify the construction by taking a suitable quotient
(see Lemma~\ref{Lem: XF/act finitary} below).
\end{Rem}

Below we will introduce an algebraic machinery for $\bbC$-algebras
that heavily makes use of congruences. To be able to use this machinery to study
$\FO$-definability, we have to provide a translation to $\bbT$-algebras
and we have to understand how congruences behave under this translation.
\begin{Def}
Let $\frakA$~be an $\bbM$-algebra.

(a)
An equivalence relation $\theta \subseteq A \times A$ is \emph{congruence} if
it induces a subalgebra of the product $\frakA \times \frakA$, that is, if
\begin{align*}
  \pi(\bbM p(u)) \mathrel\theta \pi(\bbM q(u))\,,
  \quad\text{for every } u \in \bbM\theta\,,
\end{align*}
where $p,q : A \times A \to A$ are the two projections.

(b) We denote the minimal congruence (i.e., equality) by~$\bot$ and the maximal one
(i.e., the universal relation) by~$\top$.
\end{Def}

In light of our interpretation of elements $a \in A_\xi$ as operations
$A_\emptyset^\xi \to A_\emptyset$,
we are particularly interested in congruences respecting this point of view.
\begin{Def}
Let $\frakA$~be a $\bbT$-algebra.

(a) For an equivalence relation $\theta \subseteq A_\emptyset \times A_\emptyset$
and elements $a,b \in A_\xi$, we define
\begin{align*}
  a \simeq_{\act[\theta]} b
  \quad\defiff\quad
  a(\bar c) \mathrel\theta b(\bar c)\,, \quad\text{for all } \bar c \in A_\emptyset^\xi\,.
\end{align*}
If $\theta = {=}$ is equality, we simplify the notation to~$\simeq_\act$.

(b) A congruence~$\theta$ of~$\frakA$ is \emph{saturated}
if $\theta = {\simeq_{\act[\theta_\emptyset]}}$.

(c) We call $\frakA$~\emph{reduced} if $\simeq_\act$~is equality.
\end{Def}

\begin{Lem}
A congruence~$\theta$ is saturated if, and only if, $\frakA/\theta$ is reduced.
\end{Lem}

As remarked above, $\bbC$-algebras of the form~$\bbX\frakA$ are usually not finitary.
But since we are only interested in reduced algebras, we can quotient by the
relation~$\simeq_\act$ to obtain an algebra that is both reduced and finitary.
\begin{Lem}\label{Lem: XF/act finitary}
Let\/ $\frakA$~be a\/ $\bbT$-algebra.
If\/ $\frakA$~is finitary, so is\/~$\bbX\frakA/{\simeq_\act}$.
\end{Lem}
\begin{proof}
%
$\bbX_\emptyset A = A_\emptyset$ is finite. Consequently, there are only finitely many
functions $(\bbX_\emptyset A)^\xi \to \bbX_\emptyset A$, for each fixed $\xi \in \Xi$.
Since each element $\bbX_\xi A/{\simeq_\act}$ is uniquely determined by the
function on $\bbX_\xi A$ it induces, it follows that there are only finitely many such elements.
\end{proof}

Let us quickly show that all syntactic algebras are reduced.
Therefore we can simplify the material below by working with reduced algebras only.
\begin{Lem}
Let $K \subseteq \bbT_\emptyset\Sigma$. Then\/ $\Syn(K)$ is reduced.
\end{Lem}
\begin{proof}
We can define the syntactic algebra as a quotient $\bbT\Sigma/{\approx_K}$
by the equivalence relation
\begin{align*}
  s \approx_K t \quad\defiff\quad (p[s] \in K \Leftrightarrow p[t] \in K)\,,
  \quad\text{for every context } p\,.
\end{align*}
Here a \emph{context} is a term $p \in \bbT(\Sigma + \{\Box\})$ containing
a special symbol~$\Box$ and $p[s]$ denotes the term obtained from~$p$ by replacing
every occurrence of~$\Box$ by the term~$s$.

To see that $\Syn(K)$ is reduced, consider two elements $a,b$ of arity~$n$ such that
\begin{align*}
  a(\bar c) = b(\bar c)\,, \quad\text{for all tuples } \bar c\,.
\end{align*}
Let $q : \bbT\Sigma \to \bbT\Sigma/{\approx_K} = \Syn(K)$ be the quotient map.
We fix terms $s \in q^{-1}(a)$ and $t \in q^{-1}(b)$.
To prove that $a = b$ it is sufficient to show that $s \approx_K t$.
Hence, let $p$~be a context with $p[s] \in K$.
We have to show that $p[t] \in K$.
To do so we may assume, without loss of generality, that $p$~contains exactly one occurrence
of the symbol~$\Box$. (Otherwise, we can proceed in several steps, in each one replacing a
single occurrence of~$s$ by~$t$. The claim then follows by transitivity of~$\approx_K$.)
Hence, $p = p_0(\Box(\bar r))$, for some terms $p_0$~and~$\bar r$.
Set $c_i := q(r_i)$. Then
\begin{align*}
  a(\bar c) = b(\bar c)
  \qtextq{implies}
  s(\bar r) \approx_K t(\bar r)\,.
\end{align*}
Since $\approx_K$~is a congruence, it follows that
\begin{align*}
  p[s] = p_0(s(\bar r)) \approx_K p_0(t(\bar r)) = p[t]\,.
\end{align*}
\upqed
\end{proof}
\begin{Rem}
This property is a notable difference to the theory for languages of infinite trees
where syntactic algebras do not need to be reduced.
\end{Rem}

Since we have to translate between clones and preclones we need to know how the
corresponding notions of a congruence are related.
\begin{Lem}
Let $\frakA$~be a $\bbC$-algebra and $\theta \subseteq A \times A$ an equivalence relation.
The following statements are equivalent.
\begin{enum1}
\item $\theta$~is a congruence of\/~$\frakA$.
\item $\theta$~is a congruence of\/~$\frakA|_\bbT$ satisfying
  \begin{align*}
    a \mathrel\theta b \qtextq{implies}
    {}^\sigma a \mathrel\theta {}^\sigma b\,, \quad\text{for all } a,b \in A_\xi
    \text{ and } \sigma : \xi \to \zeta\,.
  \end{align*}
\end{enum1}
\end{Lem}
\begin{proof}
(1)~$\Rightarrow$~(2) is trivial.
For the converse, suppose that $\theta$~satisfies~(2).
To show that $\theta$~is a congruence, let $u \in \bbC_\xi\theta$ and let
$s,t \in \bbC A$ be its projections to the two components.
We can write $u = {}^\sigma u'$, for some $u' \in \bbT_\zeta\theta$ and some
function $\sigma : \zeta \to \xi$.
Then $s = {}^\sigma s'$ and $t = {}^\sigma t'$, for suitable $s',t' \in \bbT_\zeta A$.
By~(2), it follows that
\begin{align*}
  \pi(s') \mathrel\theta \pi(t')
  \qtextq{and}
  \pi(s) = {}^\sigma\pi(s') \mathrel\theta {}^\sigma\pi(t') = \pi(t)\,.
\end{align*}
\upqed
\end{proof}

The next result is what is needed below to translate the algebraic machinery
from~\cite{HobbyMcKenzie88} to our current setting.
\begin{Lem}
Let $\frakA$~be a $\bbC$-algebra and $\theta \subseteq A \times A$ an equivalence relation.
The following statements are equivalent.
\begin{enum1}
\item $\theta$~is a saturated congruence of\/~$\frakA$.
\item $\theta$~is a saturated congruence of\/~$\frakA|_\bbT$.
\item $\theta = {\simeq_{\act[\delta]}}\,,\quad$%
  for some congruence~$\delta$ of\/ $\Alg(\frakA)$.
\end{enum1}
\end{Lem}
\begin{proof}
(1)~$\Rightarrow$~(2)~$\Rightarrow$~(3) is trivial.
For the remaining direction, suppose that $\theta$~satisfies~(3).
To show that $\theta$~is a congruence, let $u \in \bbC_\xi\theta$ and let
$s,s' \in \bbC A$ be its projections to the two components.
We prove that $\pi(s) \mathrel\theta \pi(s')$ by induction on~$u$.

If $u = x$ is a variable, we have $s = x = s'$, which implies that $\pi(s) = \pi(s')$.
Otherwise, $u = \langle a,a'\rangle(\bar w)$, for some terms~$\bar w$.
Then $a \mathrel\theta a'$ and $\theta = {\simeq_{\act[\delta]}}$ implies that
\begin{align*}
  a(\bar c) \mathrel\delta b(\bar c)\,, \quad\text{for all } \bar c\,.
\end{align*}
Furthermore, we have $s = a(\bar t)$ and $s' = a'(\bar t')$, where
$t_i$~and~$t'_i$ are the two projections of~$w_i$.
By inductive hypothesis, it follows that
\begin{align*}
  b_i := \pi(t_i) \mathrel\theta \pi(t'_i) =: b'_i\,, \quad\text{for all } i\,.
\end{align*}
For every tuple~$\bar c$, it therefore follows that
\begin{align*}
  \pi(s)(\bar c)
  = a(\bar b)(\bar c)
  \mathrel\delta a'(\bar b)(\bar c)
  \mathrel\delta a'(\bar b')(\bar c)
  = \pi(s')(\bar c)\,.
\end{align*}
(Here $a(\bar b)(\bar c)$ denotes the term $a(b_0(\bar c_0),\dots,b_{n-1}(\bar c_{n-1}))$,
where $\bar c_i$~is the subtuple of~$\bar c$ corresponding to the variables in~$t_i$.)
This implies that $\pi(s) \mathrel\theta \pi(t)$.
\end{proof}

\section{Minimal Algebras}   
\label{Sect:minimal}

Our main tool will be from a branch of Universal Algebra called
Tame Congruence Theory~\cite{HobbyMcKenzie88}. One of the basic results of this theory
is a classification of algebras up to the following notion of equivalence.
\begin{Def}
Two $\bbT$-algebras $\frakA$~and~$\frakB$ are \emph{polynomially equivalent} if
there exists a bijection $\varphi : A_\emptyset \to B_\emptyset$ such that
for every element $a \in A_\xi$, there exist some element $b \in B_\zeta$
and a surjective function $\sigma : \zeta \to \xi$ such that
\begin{align*}
  \varphi(a(\bar c)) = {}^\sigma b(\varphi(\bar c))\,,
  \quad\text{for all } \bar c \in A_\emptyset^\xi\,,
\end{align*}
and vice versa, for every $b \in B_\xi$, there are an element $a \in A_\zeta$ and
a surjective function $\sigma : \zeta \to \xi$ such that
\begin{align*}
  \varphi({}^\sigma a(\bar c)) = b(\varphi(\bar c))\,,
  \quad\text{for all } \bar c \in A_\emptyset^\xi\,.
\end{align*}
Here ${}^\sigma a(\bar c)$ (which is not defined for $\bbT$-algebras) is a short-hand for
\begin{align*}
  a(c_{\sigma(z_0)},\dots,c_{\sigma(z_{n-1})})\,,
\end{align*}
where $\zeta = \{z_0,\dots,z_{n-1}\}$,
\end{Def}

\begin{Lem}
Every $\bbT$-algebra~$\frakA$ is polynomially equivalent to $\frakA/{\simeq_\act}$
and to~$\bbX\frakA$.
\end{Lem}
\begin{proof}
Note that all three algebras have the same domain~$A_\emptyset$ of sort~$\emptyset$.
An element $a \in A_\xi$ of~$\frakA$ can be represented by the elements
$[a] \in \frakA/{\simeq_\act}$ and $\langle\id,a\rangle \in \bbX\frakA$.
Conversely, an $\simeq_\act$-class $[a] \in A_\xi/{\simeq_\act}$ can be represented
by the element $a$ (and the map~$\id$),
and an element $\langle\sigma,a\rangle \in \bbX_\xi A$ by the element~$a$
and the map~$\sigma$.
\end{proof}

The classification of algebras in~\cite{HobbyMcKenzie88} is based upon the following elementary
building blocks.
\begin{Def}
A $\bbC$-algebra~$\frakA$ has
\begin{itemize}
\item \emph{trivial type} (or \emph{type~T} for short) if it is polynomially
  equivalent to an algebra where every operation is constant\?;
\item \emph{unary type} (or \emph{type~U} for short) if it is non-trivial
  and polynomially equivalent to an algebra where every operation has arity at most~$1$\?;
\item \emph{affine type} (or \emph{type~A} for short) if it is polynomially
  equivalent to a vector space over a finite field\?;
\item \emph{boolean type} (or \emph{type~B} for short) if it is polynomially
  equivalent to a $2$-element boolean algebra\?;
\item \emph{lattice type} (or \emph{type~L} for short) if it is polynomially
  equivalent to a $2$-element lattice\?;
\item \emph{semilattice type} (or \emph{type~S} for short) if it is polynomially
  equivalent to $2$-element semilattice.\qedhere
\end{itemize}
\end{Def}

It turns out that we can decompose every algebra into algebras of one of these forms.
We start with the corresponding building blocks, that is, the algebras that have a type.
\begin{Def}
(a) A $\bbC$-algebra~$\frakA$ is \emph{simple} if it has exactly two saturated congruences.

(b) A $\bbC$-algebra~$\frakA$ is \emph{minimal} if it has at least two elements of
sort~$\emptyset$ and, for every $a \in A_\xi$ with $\abs{\xi} = 1$, the induced function
$A_\emptyset \to A_\emptyset$ is either constant or bijective.
\end{Def}
\begin{Thm}[\cite{HobbyMcKenzie88}]\label{Thm: 5 types of minimal algebras}
A finitary $\bbC$-algebra~$\frakA$ is minimal if, and only if, it has a type.
\end{Thm}

We are interested in subalgebras of the given algebra that are minimal.
\begin{Def}
Let $\frakA = \langle A,\pi\rangle$ be an $\bbM$-algebra where $\bbM$~is one of $\bbC$~or~$\bbT$.

(a) An $\bbM$-algebra $\frakC = \langle C,\pi_0\rangle$ is a \emph{subalgebra}
of~$\frakA$ if $C \subseteq A$ and the product~$\pi_0$ is the restriction of $\pi : \bbM A \to A$
to the set $\bbM C$.
In this case we also say that the set~$C$ \emph{induces a subalgebra} of~$\frakA$.

(b) $\frakA$~is a \emph{divisor} of a $\bbM$-algebra~$\frakB$ if
$\frakA$~is a quotient of some subalgebra of~$\frakB$.

(c) The \emph{localisation} of~$\frakA$ to a subset $C \subseteq A_\emptyset$ is
\begin{align*}
  \lsem C\rsem_\act :=
    \set{ a \in A_\xi }
        { \xi \in \Xi\,,\ a(\bar c) \in C\,, \text{ for all } \bar c \in C^\xi }\,.
\end{align*}
\upqed
\end{Def}
\begin{Rem}
Note that, when working with $\Sigma$-algebras for some algebraic signature~$\Sigma$,
the above notion of a subalgebra is stronger than the usual notion from Universal Algebra
since it also allows for the removal of operations. Hence, our notion of a subalgebra
combines the notions of a subalgebra and a reduct from Universal Algebra.
\end{Rem}

Let us check that every localisation forms a subalgebra.
\begin{Lem}
Let\/ $\frakA$~be a $\bbC$-algebra and $C \subseteq A$.
The set\/ $\lsem C\rsem_\act$ induces a subalgebra of\/~$\frakA$.
\end{Lem}
\begin{proof}
Let $t \in \bbC_\xi\lsem C\rsem_\act$. To prove that $\pi(t) \in \lsem C\rsem_\act$,
we have to show that
\begin{align*}
  \pi(t)(\bar c) \in C\,, \quad\text{for all } \bar c \in C^\xi.
\end{align*}
Hence, fix $\bar c \in C^\xi$.
Let $t'$~be the tree obtained from~$t$ by replacing each variable $x \in \xi$ by the
corresponding element~$c_x$. Then $\pi(t)(\bar c) = \pi(t')$.
Furthermore, since each operation in~$t'$ belongs to $\lsem C\rsem_\act$,
it follows by induction on~$t'$ that $\pi(t') \in C$.
\end{proof}
\begin{Rem}
It follows that $\lsem C\rsem_\act$ is the largest subalgebra of~$\frakA$ that contains
only elements of sort~$\emptyset$ which belong to~$C$.
\end{Rem}

We are particularly interested in subalgebras generated by an idempotent element as follows.
\begin{Def}
Let $\frakA$~be a $\bbT$-algebra.

(a) Each element $a \in A_\xi$ induces a function $\hat a : A_\emptyset \to A_\emptyset$ by
\begin{align*}
  \hat a(c) := a(c,\dots,c)\,.
\end{align*}

(b) An \emph{idempotent} of~$\frakA$ is an element $e \in A_\xi$ with $\xi \neq \emptyset$
satisfying
\begin{align*}
  \hat e \circ \hat e = \hat e\,.
\end{align*}

(c) For an idempotent~$e$, we set
\begin{align*}
  \lsem e\rsem := \lsem C\rsem_\act\,,
  \qtextq{where}
  C := \set{ \hat e(c) }{ c \in A_\emptyset }\,.
\end{align*}

(d) We denote the set of all idempotents $e \in A$ by~$E(\frakA)$ and
we define an order on $E(\frakA)$ by
\begin{align*}
  e \leq f \quad\defiff\quad \lsem e\rsem \subseteq \lsem f\rsem\,.
\end{align*}
\upqed
\end{Def}


We usually identify two idempotents $e$~and~$f$ if they are equivalent in this ordering, i.e.,
if $\lsem e\rsem = \lsem f\rsem$.
In particularly, note that there are only finitely many idempotents modulo this identification.
As an example, let us consider the following $\FO$-definable algebra.
\begin{Exam}
Let $\Sigma := \{a,c\}$ where $a$~is binary and $c$~a constant and
let $K \subseteq \bbT_\emptyset\Sigma$ be the language of all trees where
every leaf has an even distance from the root.
Surprisingly, it can be shown~\cite{Potthoff94,Potthoff95} that $K$~is $\FO$-definable.
The syntactic algebra $\Syn(K)$ has the following elements of sort $\xi \in \Xi$
(which we identify with functions $\{0,1,\bot\}^\xi \to \{0,1,\bot\}$).
For every $\bar u \in \{0,1\}^\xi$, we have the elements
\begin{align*}
  \bot_\xi(\bar d)   &:= \bot\,, \\
  0_{\bar u}(\bar d) &:= \begin{cases}
                           0 &\text{if } d_x = u_x \text{ for all } x\,, \\
                           \bot &\text{otherwise}\,,
                         \end{cases} \\
  1_{\bar u}(\bar d) &:= \begin{cases}
                           1 &\text{if } d_x = u_x \text{ for all } x\,, \\
                           \bot &\text{otherwise}\,,
                         \end{cases}
\end{align*}
In addition, it contains all elements generated by these.
Such elements~$a_{\bar u}$ are of the form
\begin{align*}
  a_{\bar u}(\bar d) &:= \begin{cases}
                           0 &\text{if } d_x = u_x \text{ for all } x\,, \\
                           1 &\text{if } d_x = 1 - u_x \text{ for all } x\,, \\
                           \bot &\text{otherwise}\,.
                         \end{cases}
\end{align*}
for certain $\bar u \in \{0,1\}^\xi$, but not every such tuple corresponds to an element.
Tuples~$\bar u$ where~$a_{\bar u}$ exists are, for instance,
\begin{align*}
  11\,,\quad
  100\,,\quad
  010\,,\quad
  001\,,\quad
  1111\,,\quad
  \dots
\end{align*}

There are $3$~non-trivial equivalence relations on the set~$\Syn_\emptyset(K)$.
\begin{itemize}
\item The one identifying $0$~and~$1$ is not a congruence since
  \begin{align*}
    a_{11}(1,1) = 0 \neq \bot = a_{11}(1,0)\,.
  \end{align*}
\item The one identifying $0$~and~$\bot$ is not a congruence since
  \begin{align*}
    a_{100}(1,0,0) = 0 \neq \bot = a_{100}(1,0,\bot)\,.
  \end{align*}
\item The one identifying $1$~and~$\bot$ is not a congruence since
  \begin{align*}
    a_{11}(1,1) = 0 \neq \bot = a_{11}(1,\bot)\,.
  \end{align*}
\end{itemize}
Consequently, $\Syn(K)$ is simple.

The algebra has three idempotents $E(\Syn(K)) = \{\bot_{\{x\}}, 0_0, 1_1\}$, and
the corresponding subalgebras are
\begin{align*}
  \lsem 0_0\rsem = \{0,\bot\}\,,\quad \lsem1_1\rsem = \{1,\bot\}\,,
  \qtextq{and}
  \lsem\bot_{\{x\}}\rsem = \{\bot\}\,.
\end{align*}
The first two are minimal of semilattice type, while the last one is trivial.
\end{Exam}

We conclude this section with a result of~\cite{HobbyMcKenzie88} which
generalises Theorem~\ref{Thm: 5 types of minimal algebras} to non-minimal algebras
by looking at the minimal algebras a given algebra contains.
To state the result we need to introduce some rather technical definitions.
Readers are encouraged to only skim the rest of this section and to come back later if necessary.
\begin{Def}
Let $\frakA$~be a $\bbC$-algebra and $\alpha < \beta$ congruences.

(a) We denote the $\alpha$-class of an alement $a \in A$ by $[a]_\alpha$.

(b) We call $\alpha,\beta$ a \emph{prime quotient} of~$\frakA$ if $\alpha < \beta$,
both congruences are saturated, and there is no saturated
congruence between $\alpha$~and~$\beta$.

(c) An element $a \in A_{\{x\}}$ is \emph{$\alpha\beta$-separating}
if $\hat a[\beta] \nsubseteq \alpha$.

(d) We set
\begin{align*}
  \Min_\frakA(\alpha,\beta) :=
    \biglset \lsem e\rsem \bigmset {}
             & e \in E(\frakA) \text{ a minimal $\alpha\beta$-separating} \\
             & \text{idempotent of } \frakA \bigrset\,.
\end{align*}
Elements of $\Min_\frakA(\alpha,\beta)$ are called \emph{$\alpha\beta$-minimal sets}
and the corresponding idempotents~$e$ are called \emph{$\alpha\beta$-minimal idempotents.}
For simplicity, we usually omit the subscript~$\frakA$ and simply write $\Min(\alpha,\beta)$.

(e) The algebra~$\frakA$ is \emph{$\alpha\beta$-minimal} if
$A_\emptyset \in \Min_\frakA(\alpha,\beta)$.

(f) An \emph{$\alpha\beta$-trace}
is a set of the form $[a]_\beta \cap C_\emptyset$ where $C \in \Min(\alpha,\beta)$
and $[a]_\beta \cap C_\emptyset$ intersects at least two different $\alpha$-classes.
%
%
%
\end{Def}

\begin{Def}
Let $\frakA$~be a $\bbC$-algebra, $\alpha < \beta$ two congruences of~$\frakA$, and
$X \in \{\rmT,\rmU,\rmA,\rmB,\rmL,\rmS\}$ a type. We say that $\frakA$~has
\emph{$\alpha\beta$-type~$X$} if, for every $\alpha\beta$-trace~$C$,
the divisor $\lsem C\rsem_\act/(\alpha|_C)$ is a minimal algebra of type~$X$.
\end{Def}

\begin{Thm}[\cite{HobbyMcKenzie88}]\label{Thm: characterisation of alpha,beta-minimal algebras}
Let $\alpha<\beta$ be a prime quotient of a $\bbC$-algebra~$\frakA$.
Every subalgebra $\frakC \in \Min(\alpha,\beta)$ has an $\alpha\beta$-type.
\end{Thm}

\section{Non-Definable Algebras}   
\label{Sect:non-definable}

We have shown in Theorem~\ref{Thm: FO-definable if Syn(K) FO-definable} that,
if we want to know whether a given language~$K$ is $\FO$-definable,
it is sufficient to check whether its syntactic algebra $\Syn(K)$ is $\FO$-definable.
As $\Syn(K)$ can be computed from~$K$ (for details see~\cite{Blumensath20}),
we are therefore interested in decidable characterisations of when a given algebra is
$\FO$-definable.
We start by deriving several conditions implying non-definability.
Many of the following results are already mentioned (without proof) in~\cite{BojanczykMiXX},
but note that some of the difficulties and counterexamples presented there
are circumvented by our particular technical choices regarding our algebraic framework
(like working with $\bbT$-algebras instead of $\bbC$-algebras).

We start with two composition lemmas for first-order logic over trees
that have been extracted from~\cite{Blumensath21,Potthoff92}.
\begin{Def}
For a constant $m < \omega$, two terms $s,t \in \bbC_\xi A$, and two tuples of vertices
$\bar u,\bar v$, we write
\begin{align*}
  \langle s,\bar u\rangle \equiv^m_\FO \langle t,\bar v\rangle
  \quad\defiff\quad
  &s \models \varphi(\bar u)
  \Leftrightarrow
  t \models \varphi(\bar v)\,,
  \quad\text{for all $\FO$-formulae} \\
  &\text{$\varphi(\bar x)$ of quantifier-rank at most~$m$.}
\end{align*}
\upqed
\end{Def}
\begin{Prop}[\cite{Blumensath21}]\label{Prop: composition for FO}
$\equiv^m_\FO$~is a congruence on~$\bbT\Sigma$.
\end{Prop}
\begin{Lem}[\cite{Potthoff92}]\label{Lem: adding constants to subtrees}
Let $s,t \in \bbT\Sigma$ be trees and $u,v$ vertices such that
\begin{align*}
  \langle s,u\rangle \equiv_\FO^m \langle t,v\rangle
\end{align*}
and suppose that there exists an isomorphism $\varphi : s|_u \to t|_v$.
Then
\begin{align*}
  \langle s,\bar w\rangle \equiv_\FO^m \langle t,\varphi(\bar w)\rangle\,,
  \quad\text{for every tuple } \bar w \text{ in } s|_w\,.
\end{align*}
\end{Lem}
\begin{proof}
We use a game argument (for an introduction to Ehrenfeucht-Fra\"iss\'e games see,
e.g.,~\cite{EbbinghausFlum95}).
By assumption, Duplicator has a winning strategy in the Ehrenfeucht-Fra\"iss\'e game
between $\langle s,u\rangle$ and $\langle t,v\rangle$.
We construct a strategy for Duplicator in the game between
$\langle s,\bar w\rangle$ and $\langle t,\varphi(\bar w)\rangle$ as follows.
If Spoiler chooses some vertex in the subtree attached at $u$~or~$v$,
Duplicator replies according to the isomorphism~$\varphi$.
Otherwise, Duplicator answers by using her strategy in the other game.
This strategy is clearly winning.
\end{proof}

\begin{Lem}[\cite{Potthoff92}]\label{Lem: FO-composition for trees}
Let $t \in \bbT_{\{x,x'\}}\{a,c\}$ be a tree where
$a$~is a symbol of arity greater than~$1$ and $c$~a constant symbol,
let $s,s' \in \bbT_\emptyset\Sigma$, for some alphabet~$\Sigma$,
and let $u,u' \in \dom(t)$ be two leaves.
Let $g_0$~be some function mapping each leaf of~$t$ to $s$~or~$s'$
with $g_0(u) \neq g_0(u')$,
and let $g_1$~be the function obtained from~$g_0$ by switching the values of $u$~and~$u'$.
Then
\begin{align*}
  \langle t,u\rangle \equiv_\FO^m \langle t,u'\rangle
  \text{\ \ and\ \ }
  s \equiv_\FO^m s'
  \qtextq{implies}
  t_0 \equiv_\FO^{m+1} t_1\,,
\end{align*}
where $t_i$~is the tree obtained from~$t$ by replacing each leaf~$v$ by the tree~$g(v)$.
\end{Lem}
\begin{proof}
To show that $t_0 \equiv_\FO^{m+1} t_1$ we establish the back-and-forth property.
Hence, let $v \in \dom(t_0)$.
We reply with the vertex
\begin{align*}
  v' := \begin{cases}
          u'w &\text{if } v = uw\,, \\
          uw  &\text{if } v = u'w\,, \\
          v   &\text{otherwise}\,.
        \end{cases}
\end{align*}
We claim that
\begin{align*}
  \langle t_0,v\rangle \equiv_\FO^m \langle t_1,\varphi(v)\rangle\,.
\end{align*}
For the proof we distinguish three cases.

First, suppose that $v \geq u$. By Proposition~\ref{Prop: composition for FO},
\begin{align*}
  \langle t,u\rangle \equiv_\FO^m \langle t,u'\rangle
  \text{\ \ and\ \ }
  s \equiv_\FO^m s'
  \qtextq{implies}
  \langle t_0,u\rangle \equiv_\FO^m \langle t_1,u'\rangle\,.
\end{align*}
Since $\langle t_0|_u,v\rangle \cong \langle t_1|_{u'},v'\rangle$, it follows
by Lemma~\ref{Lem: adding constants to subtrees} that
\begin{align*}
  \langle t_0,v\rangle \equiv_\FO^m \langle t_1,v'\rangle\,.
\end{align*}

If $v \geq u'$, we proceed analogously.
Finally, suppose that neither $v \geq u$ nor $v \geq u'$.
By Proposition~\ref{Prop: composition for FO},
\begin{align*}
  t|_w \equiv_\FO^m t|_w
  \qtextq{and}
  s \equiv_\FO^m s'
  \qtextq{implies}
  t_0|_w \equiv_\FO^m t_1|_w\,,
\end{align*}
for every successor~$w$ of~$v$. Furthermore, we trivially have
\begin{align*}
  t[\emptyseq,v) \equiv_\FO^m t[\emptyseq,v)
  \qtextq{and}
  \langle t(v),v\rangle \equiv_\FO^m \langle t(v),v\rangle\,.
\end{align*}
Another application of Proposition~\ref{Prop: composition for FO} therefore yields
\begin{align*}
  \langle t_0,v\rangle \equiv_\FO^m \langle t_1,v\rangle\,.
\end{align*}
\upqed
\end{proof}


In addition, we need the following observation by Potthoff about semilattices.
\begin{Lem}[Potthoff~\cite{Potthoff92}]\label{Lem: distinguishing lattice terms}
Let $\frakA$~be a $\bbT$-algebra that is a lattice
and let $a \in A_{\{x,y,u,v\}}$ be the element corresponding
to the term $(x \sqcup y) \sqcap (u \sqcup v)$.
For every $0 < n < \omega$ and every two leaves $w,w'$ of the
(tree corresponding to the) term~$a^n$ that are not siblings,
there exist constants $\bar c,\bar c' \in \{\bot,\top\}^{4^n}$ such that
\begin{itemize}
\item $a^n(\bar c) = \bot \qtextq{and} a^n(\bar c') = \top$\,,
\item $c_i = c'_i\,, \quad\text{for all~$i$ that do not correspond to $w$~or~$w'$}\,,$
\item $c_w = \bot\,,\quad c_{w'} = \top,\quad c'_w = \top,\quad c'_{w'} = \bot\,.$
\end{itemize}
\end{Lem}

Finally, we can give several different conditions implying that a given algebra is
not $\FO$-definable. None of these results is really new, or very deep\?:
condition~\textsc{(i)} is first mentioned (without proof) in~\cite{Thomas84}\?;
\textsc{(ii)}~follows directly from~\textsc{(i)}\?; \textsc{(iii)}~has already been mentioned
(without proof) in~\cite{BojanczykMiXX}\?; and \textsc{(iv)}~follows from a result
in~\cite{Potthoff92}. The only part of the following theorem that can be considered new
is the translation into the framework of $\bbT$-algebras.
\begin{Thm}\label{Thm: conditions for non-FO-definability}
If\/ $\frakA$~is a finitary\/ $\bbT$-algebra satisfying one of the following conditions,
it is not\/ $\FO$-definable.
\begin{enumi}
\item $\SG(\frakA)$ is not aperiodic.
\item $\frakA$ is polynomially equivalent to an algebra where
  every operation has arity at most~$1$ and at least one operation acts as a non-trivial
  permutation of~$A_\emptyset$.
\item $\frakA$ is polynomially equivalent to a vector space over a finite field.
\item $\frakA$ is polynomially equivalent to a lattice or a boolean algebra.
\end{enumi}
\end{Thm}
\begin{proof}
\textsc{(i)}
Since $\SG(\frakA)$ is not aperiodic,
there exists some element $a \in A_{\{x\}}$ generating a finite group.
For every $m < \omega$, there is some $n < \omega$ such that $a^n \equiv_\FO^m a^{n+1}$
(where $a^n$~denotes a path consisting of $n$~vertices labelled~$a$ and one final vertex
labelled by the variable~$x$).
Since $\pi(a^n) \neq \pi(a^{n+1})$, $\frakA$~cannot be $\FO$-definable.

\textsc{(ii)}
By assumption we can find, for every $a \in A_\xi$, some $a_0 \in A_{\{x\}}$ and $z \in \xi$
such that
\begin{align*}
  a(\bar c) = a_0(c_z)\,, \quad\text{for all } \bar c \in A_\emptyset^\xi.
\end{align*}
Furthermore, there is at least one element $a \in A_\xi$ such that the corresponding
element~$a_0$ acts as a permutation of~$A_\emptyset$.
Fixing $n > 0$ such that $a_0^n = \id$, it follows that the set
$\{a_0,a_0^2,\dots,a_0^n\}$ forms a non-trivial subgroup of~$\SG(\frakA)$.
By~\textsc{(i),} this implies that $\frakA$~is not $\FO$-definable.

\textsc{(iii)}
We use the fact that $\frakA$~is polynomially equivalent to a vector space two times\?:
first, to find an element $p \in A$ such that $p(x,\dots,x,y,\dots,y)$ corresponds to vector
addition $x + y$\?; and then to find an element $c \in A_\emptyset$ and indices
$i_0<\dots<i_{k-1}$ such that $p(x_0,\dots,x_{n-1})$ corresponds to
\begin{align*}
  x_{i_0} +\dots+ x_{i_{k-1}} + c\,.
\end{align*}
Since $p(x,\dots,x,y,\dots,y)$ corresponds to $x+y$, it follows that $c = 0$ and $k \geq 2$.
Fix some element $d \in A_\emptyset$ with $d \neq 0$ and let
$a \in A_{\{x\}}$ be the element obtained from~$p$ by replacing $x_{i_0}$~by~$x$,
$x_{i_1}$~by~$d$, and all other variables by~$0$.
Then $a$~corresponds to the operation $x + d$.
As the vector space~$\frakA$ has a positive characteristic~$q$, it follows that
this element~$a$ has order $q > 1$. In particular, $\SG(\frakA)$ contains a non-trivial
group and the claim follows by~\textsc{(i).}

\textsc{(iv)}
As above, we use the fact that $\frakA$~is polynomially equivalent to a lattice two times\?:
first, to find an element $p \in A$ such that
\begin{align*}
  p(x,\dots,x,y,\dots,y,u,\dots,u,v,\dots,v)
\end{align*}
corresponds to the lattice operation $(x \sqcup y) \sqcap (u \sqcap v)$\?;
and then to a lattice term~$t$ corresponding to $p(\bar x,\bar y,\bar u,\bar v)$.
We may assume that $t$~is is conjunctive normal form.

We claim that we can choose $p$~and~$t$ such that $t$~takes the form
\begin{align*}
  (x_i \sqcup y_j) \sqcap (u_k \sqcap v_l) \sqcap s
\end{align*}
for some indices $i,j,k.l$ and some term~$s$.
We call an argument~$\bar c$ of~$t$ \emph{diagonal} if the components of~$\bar c$
corresponding to the variables~$\bar x$ have the same value, as have the components
corresponding to~$\bar y$, those corresponding to~$\bar u$, and those corresponding to~$\bar v$.

First, note that we can simplify~$t$ as follows.
If there is some clause that contains two variables of the same type
(i.e, $x_i,x_j$, or $y_i,y_j$, or $u_i,u_j$, or $v_i,v_j$, for $i \neq j$)
we can replace one of them by~$\bot$ without affecting the value of~$t$ on diagonal arguments.

Next, note that there cannot be a clause that is disjoint from at least one of
$\bar x$~and~$\bar y$ and disjoint from at least one of $\bar u$~and~$\bar v$.
For a contradiction, suppose otherwise. By symmetry, we may assume that this clause
does not contain any variable in $\bar y\bar v$.
Let $\bar c$~be the diagonal argument to~$t$ where the variables $\bar y\bar v$ have
value~$\top$ and $\bar x\bar u$ have value~$\bot$.
Then $p(\bar c) = \top$ but $t(\bar c) = \bot$.
A~contradiction.

Finally, suppose that $t$~does not have a clause of the form $x_i \sqcup y_j$
or $u_k \sqcup v_l$. By symmetry, we may assume the former.
Then every clause contains at least one variable from $\bar u\bar v$.
Let $\bar c$~be the diagonal argument assigning $\top$~to the variables $\bar u\bar v$
and $\bot$~to $\bar x\bar y$. Then $p(\bar c) = \bot$ but $t(\bar c) = \top$.
A~contradiction.

Having established the claim, let $t'$~be the term obtained from~$t$ by
replacing all variables except for $x_i,y_j,u_k,v_l$ by the value~$\top$,
and let $a \in A_{\{x,y,u,v\}}$ be the element obtained from~$p$ in the same way.
We conclude the proof using an argument of Potthoff~\cite{Potthoff92}.
Fix a quantifier rank $m < \omega$.
For sufficiently large $n < \omega$, the tree $t_m \in \bbT\{{\sqcap},{\sqcup}\}$
corresponding to the term~$a^n$ has two leaves $w,w'$ with
\begin{align*}
  \langle t_m,w\rangle \equiv_\FO^m \langle t_m,w'\rangle\,.
\end{align*}
Let $\bar c,\bar c'$ be the constants from Lemma~\ref{Lem: distinguishing lattice terms}.
Let $s_m$~be the tree obtained from~$t_m$ by replacing every~$c_i$ that is equal to~$\bot$
by~$s_{m-1}$, and every~$c_i$ that is equal to~$\top$ by~$s'_{m-1}$.
Let $s'_m$~be the tree similarly obtained by using~$\bar c'$.
By inductive hypothesis, we have $s_{m-1} \equiv_\FO^{m-1} s'_{m-1}$.
By Lemma~\ref{Lem: FO-composition for trees}, this implies that $s_m \equiv_\FO^m s'_{m-1}$.
But $\pi(s_m) = \bot$ and $\pi(s'_{m-1}) = \top$. A~contradiction.
\end{proof}
\begin{Cor}
Let\/ $\frakA$~be a\/ $\bbT$-algebra that has a divisor\/~$\frakD$ satisfying one of the
following conditions.
\begin{enuma}
\item $\SG(\frakD)$ is not aperiodic.
\item $\frakD$ is polynomially equivalent to an algebra where
  every operation has arity at most~$1$ and at least one operation acts as a non-trivial
  permutation of~$D_\emptyset$.
\item $\frakD$ is polynomially equivalent to a vector space over a finite field.
\item $\frakD$ is polynomially equivalent to a lattice or a boolean algebra.
\end{enuma}
Then\/ $\frakA$~is not\/ $\FO$-definable.
\end{Cor}
\begin{proof}
For a contradiction, suppose that $\frakA$~is $\FO$-definable.
As the $\FO$-definable $\bbT$-algebras form a pseudo-variety (see~\cite{Blumensath21}\?;
this is not true for $\bbC$-algebras), it then follows that $\frakD$~is also $\FO$-definable.
A~contradiction to the preceding theorem.
\end{proof}

By Theorem~\ref{Thm: characterisation of alpha,beta-minimal algebras} we obtain
the following corollary.
\begin{Cor}\label{Cor: neccessary condition}
If\/ $\frakA$~is a $\FO$-definable $\bbT$-algebra, every
minimal divisor of\/~$\frakA$ is either trivial or a semilattice.
\end{Cor}
We conjecture that the converse is also true.
\begin{Conj}
A finitary\/ $\bbT$-algebra\/~$\frakA$ is\/ $\FO$-definable if, and only if,
$\SG(\frakA)$ is aperiodic
and every minimal divisor of\/~$\frakA$ is either trivial or a semilattice.
\end{Conj}
(Note that aperiodicity of $\SG(\frakA)$ follows from the other condition.
We have added it for clarity.)

\section{Definable Algebras}   
\label{Sect:definable}

It remains to establish the converse. At the moment we are only able to provide partial results.
We start with a lemma which makes it easier to check whether an algebra is $\FO$-definable.
\begin{Lem}\label{Lem: factorisation into paths}
Let $\frakA$~be a finitary $\bbT$-algebra, $C \subseteq A$ a finite set of generators,
and let $C_{\neq 1} \subseteq C$ be the set of all elements $c \in C$ of arity different
from~$1$. Then $\frakA$~is $\FO$-definable if, and only if, $\SG(\frakA)$ is aperiodic
and, for every $a \in A_\emptyset$, there is some $\FO$-formula~$\varphi_a$ such that
\begin{align*}
  \pi(t) = a \quad\iff\quad t \models \varphi_a\,,
  \quad\text{for all } t \in \bbT_\emptyset C_{\neq 1} \text{ and all } a \in A_\emptyset\,.
\end{align*}
\end{Lem}
\begin{proof}
$(\Rightarrow)$
If $\frakA$~is $\FO$-definable, the existence of the formulae~$\varphi_a$ is trivial and
the condition on $\SG(\frakA)$ follows by Theorem~\ref{Thm: conditions for non-FO-definability}.

$(\Leftarrow)$
Let $a \in A_\xi$ with an arbitrary sort~$\xi$. We have to find a formula~$\varphi_a$ such that
\begin{align*}
  \pi(t) = a \quad\iff\quad t \models \varphi_a\,,
  \quad\text{for all } t \in \bbT_\xi C\,.
\end{align*}
We construct~$\varphi_a$ by induction on~$\abs{\xi}$.

First suppose that $\xi = \emptyset$.
Let $t \in \bbT_\emptyset C$.
For every path $v_0 <\dots< v_n$ where all vertices but the last one have arity~$1$,
we can compute the product
\begin{align*}
  t(v_0)\cdot\cdots\cdot t(v_{n-1}) \cdot t(v_n)
\end{align*}
since the semigroup $\SG(\frakA)$ is aperiodic.
Let $t'$~be the tree obtained from~$t$ by replacing each such path by its product.
Then $t' \in \bbT_\emptyset C_{\neq 1}$ (w.l.o.g.\ we may assume that $C$~is closed
under left-multiplication by elements of arity~$1$) and $\pi(t') = \pi(t)$.
By assumption, we can define the product $\pi(t')$ given~$t'$.
Since the function mapping~$t$ to~$t'$ is an $\FO$-interpretation,
we can define the product~$\pi(t')$ also when given~$t$.

Next, suppose that $\xi = \{x\}$ and let $t \in \bbT_\xi C$.
For every subtree $t|_v$ without variables, we can compute $\pi(t|_v)$ as in the case above.
Replacing each of these subtrees by the value of their product, we obtain a path~$s$
with $\pi(s) = \pi(t)$. We can evaluate~$\pi(s)$ since $\SG(\frakA)$ is aperiodic.

Finally, suppose that $\abs{\xi} > 1$ and let $t \in \bbT_\xi C$.
Let $w \in \dom(t)$ be the longest common prefix of all leaves labelled by a variable,
and let $(u_y)_{y \in \eta}$ be its successors.
For each $y \in \eta$, we can evaluate the product $c_y := \pi(t|_{u_y})$ by inductive hypothesis.
Hence, we can also compute $b := \pi(t|_w)$.
Let $s(x)$~be the prefix of~$t$ where the subtree~$t|_w$ is replaced by a variable.
We can compute $a := \pi(s)$ by inductive hypothesis.
Consequently, we can also determine $\pi(t) = a(b)$.
All of this can be done in first-order logic.
\end{proof}


Let us start with the easy case of semilattices.
\begin{Lem}\label{Lem: semilattices are FO-definable}
Let $\frakA$~be a finitary $\bbC$-algebra that is polynomially equivalent to a semilattice.
Then $\frakA$~is $\FO$-definable.
\end{Lem}
\begin{proof}
By Lemma~\ref{Lem: factorisation into paths}, we have to check two conditions.
First, let us prove that $\SG(\frakA)$ is aperiodic.
In a semilattice, every non-constant unary operation $a \in A_{\{x\}}$ is of the form
$x \sqcap c$, for some $c \in A_\emptyset$.
Consequently, all elements of $\SG(\frakA)$ are idempotent.
This implies that $\SG(\frakA)$ is aperiodic.

It therefore remains to construct $\FO$-formulae~$\varphi_a$
defining $\pi^{-1}(a)$, for each $a \in A_\emptyset$.
We make use of the observation that, for a tree~$t$ in a meet-semilattice,
\begin{align*}
  u \leq v  \qtextq{implies} \pi(t|_u) \leq \pi(t|_v)\,.
\end{align*}
Also note that every element $a \in A_\xi$ corresponds to an operation
$A_\emptyset^\xi \to A_\emptyset$ of the form
\begin{align*}
  x_0 \sqcap\dots\sqcap x_{n-1}
  \qtextq{or}
  x_0 \sqcap\dots\sqcap x_{n-1} \sqcap c\,,
\end{align*}
for $x_0,\dots,x_{n-1} \in \xi$ and $c \in A_\emptyset$.
We call the set $\supp(a) := \{x_0,\dots,x_{n-1}\} \subseteq \xi$ the
\emph{support} of~$a$, and $c$~its \emph{constant term.}
We say that a vertex~$v$ of a term~$t$ is \emph{reachable} if,
for every $u < v$, we have $x \in \supp(t(u))$, where $x$~is the variable such that
$v$~belongs to the subtree attached at the $x$-successor of~$u$.

We construct the desired $\FO$-formula~$\varphi_a$ by induction on the number of elements
$b > a$. Given a tree $t \in \bbT_\emptyset A$, let $t'$~be the tree obtained from~$t$ by
\begin{itemize}
\item replacing every subtree whose product is greater than~$a$ by a leaf
  of that value and
\item merging these leaves into their parent vertex.
\end{itemize}
By inductive hypothesis, there exists an $\FO$-interpretation mapping~$t$ to~$t'$.
Given~$t$, the formula~$\varphi_a$ states the following properties of the modified tree~$t'$.
\begin{itemize}
\item For every reachable vertex~$v$ such that the operation $t(v)$ has a constant term~$c$,
  we have $c \geq a$.
\item Every reachable leaf is labelled by~$a$.
\item Every reachable internal vertex is labelled by an operation mapping
  the tuple $a\dots a$ to~$a$.\qedhere
\end{itemize}
\end{proof}

Our next goal is to show that, in order to define the product of a given term,
it is sufficient to be able to determine to which minimal sets it belongs.
One technical issue we have to deal with is that the notion of an idempotent
is defined for $\bbC$-algebras, but here we deal with $\bbT$-algebras.
This means we have to use elements of the form $e(x,\dots,x)$ where the arity of~$e$
can be larger than~$1$. Such elements do not belong to the given algebra~$\frakA$,
but to $\bbX\frakA$.
\begin{Def}
Let $\frakA$~be a $\bbT$-algebra and $\theta$~a congruence of~$\frakA$.
A term $t \in \bbT\bbX A$ is in \emph{$\theta$-idempotent normal form} if,
for every $v \in \dom(t)$, there exist elements $e,a \in \bbX A$
and a function~$\sigma$ such that
$t(v) = \langle \sigma,ea\rangle$ and $e$~is a $\bot\theta$-minimal idempotent.
\end{Def}

\begin{Lem}\label{Lem: evaluating terms in idempotent normal form}
Let\/ $\frakA$~be a finitary\/ $\bbT$-algebra such that every minimal divisor
is of semilattice type, $C$~a finite set of generators, and $\theta$~a congruence of\/~$\frakA$.
There exists a family $(\varphi_c)_{c \in A_\emptyset}$ of\/ $\FO$-formulae such that,
for every term $t \in \bbT_\emptyset\bbX C$ in $\theta$-idempotent normal form,
\begin{align*}
  t \models \varphi_c \quad\iff\quad \pi(t) = c\,.
\end{align*}
\end{Lem}
\begin{proof}
Fix a tree $t \in \bbT_\emptyset\bbX C$
which we represent as a $C$-labelled acyclic directed graph~$\frakG$ defined as follows.
The vertices of~$\frakG$ are the same as those of~$t$.
For each vertex~$v$ with label $t(v) = \langle\sigma,a\rangle$,
we label~$v$ in~$\frakG$ by the element~$a$ and, for every variable~$z$ of~$a$,
there is an outgoing $z$-labelled edge to the $\sigma(z)$-successor of~$v$ in~$t$.
It follows that the unravelling of~$\frakG$ produces a term $\hat t \in \bbT_\emptyset C$
with the same value as~$t$.
Note that the graph~$\frakG$ is $\FO$-interpertable in~$t$.
(Although its unravelling~$\hat t$ is not.)

Let $\varepsilon$~be a function mapping each vertex~$v$ of~$t$ to some
$\bot\theta$-minimal idempotent~$e$ such that $t(v) \in eA$.
We call a factor $\hat t[u,\bar v)$ of~$\hat t$ an \emph{$e$-factor} if
$\varepsilon(u) = e$ and $\varepsilon(v_x) = e$, for all~$x$.
An $e$-factor is \emph{primitive} if $\varepsilon(w) \neq e$, for all
$w \in [u,\bar v) \setminus \{u\}$.

Note that, if $\hat t[u,\bar v)$ is an $e$-factor, we have $\pi(\hat t[u,\bar v)) \in eA$.
Since the minimal set~$eA_\emptyset$ has semilattice type,
it follows that the element $\pi(\hat t[u,\bar v))$ corresponds to a semilattice term
of the form
\begin{align*}
  \top
  \qtextq{or}
  \bot
  \qtextq{or}
  x_{j_0} \sqcap\dots\sqcap x_{j_{n-1}}\,,
  \quad\text{for some variables}\,.
\end{align*}

Given an idempotent~$e$ and a set~$M$ of idempotents, we will construct formulae
$\vartheta_\bot(x)$, $\vartheta_\top(x)$, $\psi(x,y)$, $\hat\vartheta_\bot(x)$,
$\hat\vartheta_\top(x)$, and $\hat\psi(x,y)$ with the following properties.
\begin{enumi}
\item For every \emph{primitive} $e$-factor $\hat t[u,\bar v)$ of~$\hat t$ with
  \begin{align*}
    \set{ \varepsilon(w) }{ w \in [u,\bar v) } \subseteq M\,,
  \end{align*}
  we have
  \begin{alignat*}{-1}
    \frakG &\models \vartheta_\bot(u')
      &&\quad\iff\quad &&\pi(\hat t[u,\bar v)) = \bot\,, \\
    \frakG &\models \vartheta_\top(u')
      &&\quad\iff\quad &&\pi(\hat t[u,\bar v)) = \top\,, \\
    \frakG &\models \psi(u',w')
      &&\quad\iff\quad
      &&w = v_i \text{ and } \pi(\hat t[u,\bar v)) = x_{j_0} \sqcap\cdots\sqcap x_{j_{n-1}} \\
      &&&&&\text{contains the variable } x_i\,,
  \end{alignat*}
  where $u'$~and~$w'$ are the vertices of~$\frakG$ corresponding to the vertices $u$~and~$w$
  of~$\hat t$.
\item For every \emph{maximal} $e$-factor $\hat t[u,\bar v)$ of~$\hat t$ with
  \begin{align*}
    \set{ \varepsilon(w) }{ w \in [u,\bar v) } \subseteq M\,,
  \end{align*}
  we have
  \begin{alignat*}{-1}
    \frakG &\models \hat\vartheta_\bot(u')
      &&\quad\iff\quad &&\pi(\hat t[u,\bar v)) = \bot\,, \\
    \frakG &\models \hat\vartheta_\top(u')
      &&\quad\iff\quad &&\pi(\hat t[u,\bar v)) = \top\,, \\
    \frakG &\models \hat\psi(u',w')
      &&\quad\iff\quad
      &&w = v_i \text{ and } \pi(\hat t[u,\bar v)) = x_{j_0} \sqcap\cdots\sqcap x_{j_{n-1}} \\
      &&&&&\text{contains the variable } x_i\,,
  \end{alignat*}
  where $u'$~and~$w'$ are the vertices of~$\frakG$ corresponding to the vertices $u$~and~$w$
  of~$\hat t$.
\end{enumi}
We proceed by induction on the size of~$M$.
Since the above conditions are bisimulation-invariant, it does not matter whether
the formulae we construct are evaluated in the graph~$\frakG$ or its unravelling~$\hat t$.
For simplicity, we will therefore construct formulae for~$\hat t$.

First note that, given a vertex $u \in \dom(t)$, we can use the labelling~$\varepsilon$
(which is definable given~$\frakG$) to define a set of vertices~$\bar v$ such that
$\hat t[u,\bar v)$ is a primitive/maximal $\varepsilon(u)$-factor (if such a set exists).

\textsc{(i)} Given a primitive $e$-factor $\hat t[u,\bar v)$, let
$s$~be the tree obtained from~$\hat t[u,\bar v)$ by evaluating every maximal $f$-factor
with $f \in M \setminus \{e\}$ and let $\eta$~be the corresponding labelling of~$s$
by idempotents induced by~$\varepsilon$.
Note that $s$~is not an element of $\bbT\bbX C$ but a `mixed-term' which is labelled
not only by elements of~$\bbX C$ but also by semilattice operations $\bot,\top,{\sqcap}$
(each associated with some minimal set).
Furthermore, the successors of $\sqcap$-labelled vertices are annotated as either
`relevant' or `irrelevant' depending on whether or not the operation depends on this successor.
We can use the inductive hypothesis to interpret~$s$ in $\hat t[u,\bar v)$.
We simplify~$s$ in several steps as follows.
\begin{enum1}
\item For every vertex~$v$ labelled by $\bot$~or~$\top$, we replace the attached subtree
  by a leaf labelled by the corresponding element of $\eta(v)\cdot A_\emptyset$.
\item For every successor~$v$ of a $\sqcap$-labelled vertex that is labelled by~$\top$,
  we delete the subtree attachted at~$v$.
\item If some $\sqcap$-labelled vertex~$v$ has a successor labelled~$\bot$,
  we replace the subtree attached at~$v$ by a leaf with value~$\bot$.
\end{enum1}
Let $s'$~be the resulting term.
Note that, by construction of~$s$, every branch of~$s$ contains at most two vertices labelled
with the same idempotent~$f$. Hence, the height of~$s$ is bounded by $2\cdot\abs{M}$.
Furthermore, every subtree~$s|_v$ that does not contain variables forms an $\eta(v)$-factor
of~$\hat t$ which, by construction of~$s$, implies that each such subtree consists of a single leaf.
Finally, by construction of~$s'$, the number of successors of a $\sqcap$-labelled vertex~$v$
of~$s'$ is bounded by the number of variables in the subtree $s'|_v$.
Therefore, the number of vertices of~$s'$ is bounded and
there are only finitely many possibilities of such terms.
Consequently, the operation associated with such a term can be defined by suitable
$\FO$-formulae $\vartheta_\bot(x)$, $\vartheta_\top(x)$, $\psi(x,y)$ that simply
enumerate all relevant cases.

\textsc{(ii)} It remains to consider the case where $\hat t[u,\bar v)$ is a maximal $e$-factor.
Let $H := \varepsilon^{-1}(e) \cap [u,\bar v)$ be the set of vertices with labelling~$e$.
Let us call a vertex $w \in H$ \emph{reachable} if we have
\begin{align*}
  \hat t \models \psi(w_i,w_{i+1})\,, \quad\text{for all } i < m\,,
\end{align*}
where $w_0,\dots,w_m$ is the maximal chain in~$H$ with $w_m = w$.
It follows that $\hat\vartheta_\bot(u)$ should hold if
\begin{align*}
  \hat t \models \vartheta_\bot(w)\,, \quad\text{for some reachable } w \in [u,\bar v)\,,
\end{align*}
$\hat\psi(u,v_i)$ should hold if $v_i$~is reachable and $\hat\vartheta_\bot(u)$ does not
hold, and $\hat\vartheta_\top(u)$ should hold if $\hat\vartheta_\bot(u)$ does not hold
and no $v_i$~is reachable.
Each of these conditions can be expressed in $\FO$.

To conclude the proof, note that $\dom(\hat t)$ is a maximal $e$-factor
where $e := \varepsilon(\emptyseq)$.
Hence,
\begin{align*}
  \pi(t) =
  \pi(\hat t) = \begin{cases}
             \top &\text{if } \hat t \models \hat\vartheta_\top(\emptyseq)\,, \\
             \bot &\text{if } \hat t \models \hat\vartheta_\bot(\emptyseq)\,.
           \end{cases}
\end{align*}
\upqed
\end{proof}

It follows that an algebra is $\FO$-definable if we can define the minimal sets.
\begin{Cor}\label{Cor: sufficient condition 1}
Let\/ $\frakA$~be a finitary\/ $\bbT$-algebra such that every minimal divisor
is of semilattice type, $C$~a finite set of generators, and $\theta$~a congruence of\/~$\frakA$.
If there exists a family $(\varphi_D)_{D \in \Min(\bot,\theta)}$ of $\FO$-formulae such that
\begin{align*}
  t &\models \Lor_D \varphi_D\,,
    \quad\text{for every } t \in \bbT_\emptyset C\,, \\
  t &\models \varphi_D \qtextq{implies} \pi(t) \in D\,,
    \quad\text{for all } t \in \bbT_\emptyset C
    \text{ and } D \in \Min(\bot,\theta) \,,
\end{align*}
then $\frakA$~is $\FO$-definable.
\end{Cor}
\begin{proof}
For every $v \in \dom(t)$, we fix some set~$D_v$ with $t|_v \models \varphi_{D_v}$ and
an idempotent~$e_v$ such that $D_v = e_vA_\emptyset$.
Let $s$~be the tree obtained from~$t$ by replacing each label~$t(v)$ by~$e_vt(v)$.
Note that $e_v = a(x,\dots,x)$ for some element whose arity might be larger than~$1$.
Consequently, we have to go to the algebra $\bbX\frakA$ to form the product $e_vt(v)$.
Therefore, the tree~$s$ belongs to $\bbT_\emptyset\bbX C'$ where
$C' := \set{ ec }{ c \in C,\ e \in E(\frakA) }$.
Since $s$~is in idempotent normal form, we can compute its product~$\pi(s)$ using
Lemma~\ref{Lem: evaluating terms in idempotent normal form}.
It therefore remains to show that $\pi(s) = \pi(t)$.
We do so by induction on the number of vertices.
Let $v$~be the root of~$s$ and $u_0,\dots,u_{n-1}$ be sequence of its successors.
By inductive hypothesis, we have $\pi(s|_{u_i}) = \pi(t|_{u_i})$.
Therefore,
\begin{align*}
  \pi(s)
  &= s(v)\bigl(\pi(s|_{u_0}),\dots,\pi(s|_{u_{n-1}})\bigr) \\
  &= s(v)\bigl(\pi(t|_{u_0}),\dots,\pi(t|_{u_{n-1}})\bigr) \\
  &= e_vt(v)\bigl(\pi(t|_{u_0}),\dots,\pi(t|_{u_{n-1}})\bigr) \\
  &= e_v\pi(t) \\
  &= \pi(t)\,,
\end{align*}
where the last step follows form the fact that $\pi(t) \in e_vA_\emptyset$.
\end{proof}

It follows that, in order to compute the value of a product in $\FO$, we only have to
determine which minimal sets it belongs to. Unfortunately, this seems to be as difficult
as the original question. Nevertheless, we have included the above result since the
construction in its proof seems to be useful for the full result.

Next, let us prove a weaker result where, instead of the minimal sets, we assume
that every simple divisor is a semilattice. (As the example at the beginning of
this section shows, this is indeed a stronger assumption.)
We start with a technical observation.
\begin{Lem}\label{Lem: subalgebra covering a term}
Let\/ $\frakA$~be a finitary\/ $\bbT$-algebra, $\theta$~a congruence of\/~$\frakA$, and
$D$~a $\theta$-class.
Every tree $t \in \bbT_\emptyset A$ satisfying
\begin{align*}
  \pi(t|_v) \in D\,, \quad\text{for all } v \in \dom(t)\,,
\end{align*}
belongs to $\bbT_\emptyset\lsem D_\emptyset\rsem_\act$.
\end{Lem}
\begin{proof}
We have to show that every element of the form $a = t(v)$ belongs to
$\lsem D_\emptyset\rsem_\act$. Fix a vertex $v \in \dom(t)$ and set $a := t(v)$.
For $x \in \xi$, let $u_x$~be the $x$-successor of~$v$ and set $c_x := \pi(t|_{u_x})$.
Since $\theta$~is a congruence, $\bar c \in D^\xi$ and $a(\bar c) = \pi(t|_v) \in D$
implies that $a(\bar c') \in D$, for all $\bar c' \in D^\xi$.
\end{proof}

\begin{Prop}\label{Prop: conditions for FO-definability}
Let\/ $\frakA$~be a finitary\/ $\bbT$-algebra such that\/ $\SG(\frakA)$ is aperiodic and
every simple divisor\/~$\frakD$ of\/~$\frakA$ is a semilattice.
Then\/ $\frakA$~is\/ $\FO$-definable.
\end{Prop}
\begin{proof}
The proof is analogous to that of Lemma~\ref{Lem: evaluating terms in idempotent normal form}.
Let $C \subseteq A$ be a finite set of generators, fix a sort $\xi \in \Xi$,
and let $\Delta \subseteq \Xi$ be a finite set of sorts such that $C \subseteq A|_\Delta$
and $\xi \in \Delta$.
By Lemma~\ref{Lem: factorisation into paths}, it is sufficient to
construct formulae defining the product for trees $t \in \bbT_\emptyset C$.
We proceed by induction on~$\abs{A|_\Delta}$,
distinguishing three cases based on the structure of the congruence lattice of~$\frakA$.

Every algebra~$\frakA$ has two trivial congruences\?: equality~$\bot$ and the relation~$\top$
where all elements (of the same sort) are equivalent.
If these two congruences coincide, $\frakA$~has a single element of every sort and
it is trivially $\FO$-definable.

Next, suppose that there are two non-trivial congruences $\theta,\theta'$ of~$\frakA$ with
$\theta \cap \theta' = \bot$.
Given a tree $t \in \bbT_\emptyset C$ we use the inductive hypothesis to compute
the product of~$t$ in the algebras $\frakA/\theta$~and~$\frakA/\theta'$.
Since $\theta \cap \theta' = \bot$ it follows that these two values uniquely determine $\pi(t)$.

It therefore remains to consider the case where there is a unique minimal congruence
$\theta > \bot$ (which might be equal to~$\top$).
We proceed similarly as in the proof of
Lemma~\ref{Lem: evaluating terms in idempotent normal form}.
Fix $t \in \bbT_\emptyset C$
and let $\rho$~be a function mapping each vertex~$v$ of~$t$ to the class~$[\pi(t|_v)]_\theta$.
Note that $\rho$~is definable since, by inductive hypothesis, the algebra $\frakA/\theta$~is
$\FO$-definable. We call a factor $t[u,\bar v)$ of~$t$ a \emph{$D$-factor} if
$\rho(u) = D$ and $\rho(v_x) = D$, for all~$x$.
A $D$-factor is \emph{primitive} if $\rho(w) \neq D$, for all
$w \in [u,\bar v) \setminus \{u\}$.

Note that, by Lemma~\ref{Lem: subalgebra covering a term}, we have
\begin{align*}
  \pi(t[u,\bar v)) \in \lsem D\rsem_\act\,,
  \quad\text{for every $D$-factor } [u,\bar v)\,.
\end{align*}
Since $\lsem D\rsem_\act$~is polynomially equivalent to a semilattice,
it follows that the element $\pi(t[u,\bar v))$ corresponds to a semilattice term of the form
\begin{align*}
  \top
  \qtextq{or}
  \bot
  \qtextq{or}
  x_{j_0} \sqcap\dots\sqcap x_{j_{n-1}}\,,
  \quad\text{for some variables}\,.
\end{align*}

Given a class~$D$ and a set~$M$ of $\theta$-classes,
we will construct formulae $\vartheta_\bot(x)$, $\vartheta_\top(x)$, $\psi(x,y)$,
$\hat\vartheta_\bot(x)$, $\hat\vartheta_\top(x)$, and $\hat\psi(x,y)$ with
the following properties.
\begin{enumi}
\item For every \emph{primitive} $D$-factor $t[u,\bar v)$ of~$t$ with
  \begin{align*}
    \set{ \rho(w) }{ w \in [u,\bar v) } \subseteq M\,,
  \end{align*}
  we have
  \begin{alignat*}{-1}
    t &\models \vartheta_\bot(u) &&\quad\iff\quad &&\pi(t[u,\bar v)) = \bot\,, \\
    t &\models \vartheta_\top(u) &&\quad\iff\quad &&\pi(t[u,\bar v)) = \top\,, \\
    t &\models \psi(u,w)         &&\quad\iff\quad
                                 &&w = v_i \text{ and } \pi(t[u,\bar v)) = x_{j_0} \sqcap\cdots\sqcap x_{j_{n-1}} \\
                                 &&&&&\text{contains the variable } x_i\,.
  \end{alignat*}
\item For every \emph{maximal} $D$-factor $t[u,\bar v)$ of~$t$ with
  \begin{align*}
    \set{ \rho(w) }{ w \in [u,\bar v) } \subseteq M\,,
  \end{align*}
  we have
  \begin{alignat*}{-1}
    t &\models \hat\vartheta_\bot(u) &&\quad\iff\quad &&\pi(t[u,\bar v)) = \bot\,, \\
    t &\models \hat\vartheta_\top(u) &&\quad\iff\quad &&\pi(t[u,\bar v)) = \top\,, \\
    t &\models \hat\psi(u,w)         &&\quad\iff\quad
                                     &&w = v_i \text{ and } \pi(t[u,\bar v)) = x_{j_0} \sqcap\cdots\sqcap x_{j_{n-1}} \\
                                     &&&&&\text{contains the variable } x_i\,.
  \end{alignat*}
\end{enumi}
We proceed by induction on the size of~$M$.

First note that, given a vertex $u \in \dom(t)$, we can use the labelling~$\rho$
to define a set of vertices~$\bar v$ such that $t[u,\bar v)$
is a primitive/maximal $\rho(u)$-factor (if such a set exists).

\textsc{(i)} Given a primitive $D$-factor $[u,\bar v)$, let
$s$~be the tree obtained from~$t[u,\bar v)$ by evaluating every maximal $E$-factor
with $E \in M \setminus \{D\}$ and let $\eta$~be the corresponding labelling of~$s$
by $\theta$-classes induced by~$\rho$.
Note that $s$~is not an element of $\bbT C$ but a `mixed-term' which is labelled
not only by elements of~$C$ but also by semilattice operations $\bot,\top,{\sqcap}$
(each associated with some minimal set).
Furthermore, the successors of $\sqcap$-labelled vertices are annotated as either
`relevant' or `irrelevant' depending on whether or not the operation depends on this
successor. We can use the inductive hypothesis to interpret~$s$ in $t[u,\bar v)$.
We simplify~$s$ in several steps as follows.
\begin{enum1}
\item For every vertex~$v$ labelled by $\bot$~or~$\top$, we replace the attached subtree
  by a leaf labelled by the corresponding element of $\eta(v)\cdot A_\emptyset$.
\item For every successor~$v$ of a $\sqcap$-labelled vertex that is labelled by~$\top$,
  we delete the subtree attachted at~$v$.
\item If some $\sqcap$-labelled vertex~$v$ has a successor labelled~$\bot$,
  we replace the subtree attached at~$v$ by a leaf with value~$\bot$.
\end{enum1}
Let $s'$~be the resulting term.
Note that, by construction of~$s$, every branch of~$s$ contains at most two vertices labelled
with the same class~$E$. Hence, the height of~$s$ is bounded by $2\cdot\abs{M}$.
Furthermore, every subtree~$s|_v$ that does not contain variables forms an $\eta(v)$-factor
of~$t$ which, by construction of~$s$, implies that each such subtree consists of a single leaf.
Finally, by construction of~$s'$, the number of successors of a $\sqcap$-labelled vertex~$v$
of~$s'$ is bounded by the number of variables in the subtree $s'|_v$.
Consequently, the number of vertices of~$s'$ is bounded and
there are only finitely many possibilities of such terms.
Consequently, the operation associated with such a term can be defined by suitable
$\FO$-formulae $\vartheta_\bot(x)$, $\vartheta_\top(x)$, $\psi(x,y)$ that simply
enumerate all relevant cases.

\textsc{(ii)} It remains to consider the case where $t[u,\bar v)$ is a maximal $D$-factor.
Let $H := \rho^{-1}(D) \cap [u,\bar v)$ be the set of vertices with labelling~$D$.
Let us call a vertex $w \in H$ \emph{reachable} if we have
\begin{align*}
  t \models \psi(w_i,w_{i+1})\,, \quad\text{for all } i < m\,,
\end{align*}
where $w_0,\dots,w_m$ is the maximal chain in~$H$ with $w_m = w$.
It follows that $\hat\vartheta_\bot(u)$ should hold if
\begin{align*}
  t \models \vartheta_\bot(w)\,, \quad\text{for some reachable } w \in [u,\bar v)\,,
\end{align*}
$\hat\psi(u,v_i)$ should hold if $v_i$~is reachable and $\hat\vartheta_\bot(u)$ does not
hold, and $\hat\vartheta_\top(u)$ should hold if $\hat\vartheta_\bot(u)$ does not hold
and no $v_i$~is reachable.
Each of these conditions can be expressed in $\FO$.

To conclude the proof, note that $\dom(t)$ is a maximal $D$-factor
where $D := \rho(\emptyseq)$.
Hence,
\begin{align*}
  \pi(t) = \begin{cases}
             \top &\text{if } t \models \hat\vartheta_\top(\emptyseq)\,, \\
             \bot &\text{if } t \models \hat\vartheta_\bot(\emptyseq)\,.
           \end{cases}
\end{align*}
\upqed
\end{proof}

In light of the above two results, it remains to settle the following question.
\begin{Open}
Given a reduced simple $\bbT$-algebra~$\frakA$ where every $\bot\top$-minimal set has
semilattice type. Does there exist a family of $\FO$-formulae that, given a term
$t \in \bbT A$, computes (some of) the $\bot\top$-minimal sets containing $\pi(t)$\??
\end{Open}

\section{Conclusion}   

We have derived several necessary and sufficient conditions on when a tree language~$K$
is first-order definable. We have seen in Corollary~\ref{Cor: neccessary condition}
that, if $K$~is $\FO$-definable, every minimal divisor of its syntactic algebra
$\Syn(K)$ must be either trivial or (polynomially equivalent to) a $2$-element semilattice.
We could only establish the converse using additional assumptions\?:
Corollary~\ref{Cor: sufficient condition 1} assumes that the minimal sets of $\Syn(K)$
are definable, while Proposition~\ref{Prop: conditions for FO-definability} assumes
that all simple subalgebras are semilattices.

What is missing for a full characterisation of first-order logic\??
In light of Theorem~\ref{Thm: FO = chain + aperiodic}, there seem to be three conditions
for $\FO$-definability\?: \textsc{(i)}~regularity, \textsc{(ii)}~aperiodicity, and
\textsc{(iii)}~the fact that each $\FO$-formula only speaks about a bounded number of
branches of the tree. In this article we have focussed on aperiodicity.
The condition in Corollary~\ref{Cor: neccessary condition} seems to capture this part
of the characterisations to a sufficient degree.
For further progress, it seems the focus should be on~\textsc{(iii),} that is,
to try to find a characterisation for definability in chain logic.

{\small\raggedright
\bibliographystyle{siam}
\bibliography{Tame-submitted}}

\end{document}